\documentclass[aip,pop,reprint]{revtex4-1}

\usepackage{amsmath,
,amsthm
,xspace
,graphicx
,subfig
}
\begin{document}
\newcommand{\D}{\ensuremath{\Delta}\xspace}
\renewcommand{\d}{\ensuremath{\delta}\xspace}
\newcommand{\dc}{\ensuremath{\delta_c}\xspace}
\newcommand{\eps}{\ensuremath{\varepsilon}\xspace}
\newcommand{\n}{\ensuremath{\nu}\xspace}
\newcommand{\sig}{\ensuremath{\sigma}\xspace}

\renewcommand{\bar}[1]{\ensuremath{\overline{#1}}}

\newcommand{\ra}{\ensuremath{\rightarrow}}
\newcommand{\Ra}{\ensuremath{\Rightarrow}}
\newcommand{\LRa}{\ensuremath{\Leftrightarrow}}

\newcommand{\half}{\ensuremath{\frac{1}{2}}}
\newcommand{\bT}{\ensuremath{\mathbf{T}}}
\newcommand{\bR}{\ensuremath{\mathbf{R}}}
\newcommand{\mT}{\ensuremath{\leftidx{^\pm}{\overline{T}}{}}}
\newcommand{\mTp}{\ensuremath{\leftidx{^+}{\overline{T}}{}}}
\newcommand{\mTm}{\ensuremath{\leftidx{^-}{\overline{T}}{}}}
\newcommand{\mmT}{\ensuremath{\leftidx{^3}{\overline{T}}{}}}

\newcommand{\Bc}{Boundary condition\xspace}
\newcommand{\Bcs}{Boundary conditions\xspace}
\newcommand{\bc}{boundary condition\xspace}
\newcommand{\bcs}{boundary conditions\xspace}
\newcommand{\Ca}{Cellular automaton\xspace}
\newcommand{\Cas}{Cellular automata\xspace}
\newcommand{\ca}{cellular automaton\xspace}
\newcommand{\cas}{cellular automata\xspace}
\newcommand{\ea}{\textit{et al.}\xspace}
\newcommand{\etc}{\textit{etc.}\xspace}
\newcommand{\ie}{\textit{i.e.}\xspace}
\newcommand{\circa}{\textit{circa}}
\newcommand{\soc}{\textsc{Soc}\xspace}
\newcommand{\fortran}{\textsc{Fortran}\xspace}
\newcommand{\ttilde}{\ensuremath{\sim}\xspace}
\newcommand{\p}{\ensuremath{\partial}\xspace}

\newcommand{\beq}{\begin{equation}}
\newcommand{\eeq}{\end{equation}}

\title{A self-organized criticality model for ion temperature gradient (ITG) 
mode driven turbulence in confined plasma}

\author{H.\ Isliker}
\author{Th.\ Pisokas}
\affiliation{Section of Astrophysics, Astronomy and Mechanics, 
Department of Physics, Aristotle University of Thessaloniki, 
Association Euratom - Hellenic Republic, GR-54124 Thessaloniki, Greece}
\author{D.\ Strintzi}
\affiliation{National Technical University of Athens, 
Association Euratom - Hellenic Republic, GR-15773 Athens, Greece}
\author{L.\ Vlahos}
\affiliation{Section of Astrophysics, Astronomy and Mechanics, 
Department of Physics, Aristotle University of Thessaloniki, 
Association Euratom - Hellenic Republic, GR-54124 Thessaloniki, Greece}

\date{\today}

\begin{abstract}
A new Self-Organized Criticality (\soc) model is introduced in the form of a 
Cellular Automaton (CA) for ion temperature gradient (ITG) mode driven 
turbulence in fusion plasmas.
Main characteristics of the model are that it is constructed
in terms of  
the actual physical variable, the ion temperature, 
and that the temporal evolution of the CA,
which necessarily is in the form of rules, mimics actual 
physical processes as they are considered to be active in
the system, i.e.\ 
a heating process 
and a local diffusive process that sets on  if a threshold in the normalized 
ion temperature gradient $R/L_T$ is exceeded.
The model reaches the \soc state and yields
ion temperature profiles of exponential shape, 
which 
exhibit very high stiffness, in that 
they basically are independent of the loading pattern applied.
This implies that there is anomalous 
heat transport 
present in the system, despite the fact that diffusion at the local 
level
is imposed to be of a normal kind.
The distributions of the heat fluxes in the system and of 
the heat out-fluxes 
are of power-law shape.
The basic properties of the model are in 
good qualitative agreement 
with experimental results. 
\end{abstract}



\maketitle

\section{Introduction}

Self-organized criticality (\soc) is a possible state of complex, spatially 
extended systems that are systematically driven and that
have mechanisms to develop local instabilities and to relax them. 
\soc is characterized by intermittent transport events that range from very 
small size up to system size, so called avalanches, by power-law distributions
of variables characterizing the transport, and by power-spectra of the
dissipated energy that are of power-law shape. 
\soc models are usually constructed in the form of Cellular 
Automata (CA), i.e.\ by using a discrete grid and rules 
for the evolution of the system, with proto-type the sand-pile
model of Ref.\ \onlinecite{Bak87}.

In inquiries on confined plasmas and the related transport phenomena,
evidence has been collected that the plasma 
might well be in the state
of \soc, e.g.\ 
the fluctuations in
density, potential, particle flux, and electron temperature 
show avalanche-like characteristics, such as 
intermittency and power-spectra of power-law shape,
and the probability distribution of the particle flux displays a power-law
(e.g.\ Refs.\ \onlinecite{Pedrosa99}, \onlinecite{Rhodes99}, \onlinecite{Politzer00}, and see the 
recent discussion in Ref.\ \onlinecite{Sattin06}). 
Also fluid simulations have been found to exhibit \soc features,
such as the occurrence of avalanches and the appearance of
frequency spectra of power-law shape (e.g.\ Refs.\ \onlinecite{Garbet98}, 
\onlinecite{Sarazin98}). 
A discrepancy between \soc models and experimental results 
had appeared concerning 
distributions of waiting-times 
between bursts in density-fluctuation and particle-flux times-series, 
the distributions determined from experimental time-series
were found to be of power-law shape  \cite{Carbone}, 
whereas the earlier \soc models, close in form to the model of 
Ref.\ \onlinecite{Bak87},
yielded distributions of exponential shape.
It was though realized later that spatial correlations in the 
loading process can lead to power-law distributed waiting-times
\cite{Sanch02,Sattin06,Baiesi06}, 
and Ref.\ \onlinecite{Paczuski05} found that, depending on the 
threshold applied in the burst detection, the waiting time distributions 
can turn from exponential to power-law shape, even in the case of the
classical \soc model of Ref.\ \onlinecite{Bak87}.
Waiting time statistics can therefore not be considered as
an appropriate tool to test for \soc behavior in physical systems.

Several \soc models in the form of CA have been suggested for fusion plasmas
that address different aspects of turbulent transport and that are able
to reproduce a number of observed phenomena, including
the transport suppressing role of sheared poloidal flows 
\cite{Newman96}, 
the scaling of transport characteristics with system size
\cite{Carr02},
the power-law shape of waiting time distributions
\cite{Sanch02,Sanch02b,Sattin06},  
the occurrence of anomalous, super-diffusive transport events
\cite{Carr02b},
the appearance of enhanced confinement, edge pedestals, edge localized modes, 
and the L- to H-mode transition
\cite{Chapman01,Gruzinov02,Gruzinov03,Sanchez03}, 
and the non-diffusive energy transport observed in off-axis heating 
experiments
\cite{March04}.

These \soc models are all CA models of the sand-pile type, and they basically 
are variants of the original \soc model of Ref.\ \onlinecite{Bak87}
and its generalizations by Ref.\ \onlinecite{Kadanoff89}, which also allows
non-local relaxation of the instabilities, and  
Ref.\ \onlinecite{Hwa92}, which introduces the running sand-pile model
that is continuously loaded. The basic elements of the models thus are 
sand-grains, height and height-differences of sand-columns, with a usually
vague identification of these system variables with the physical
variables such as density, energy density, or temperature, 
and with an alike vague association of the system dynamics with
actual physical processes. 
(A different approach is followed in
Ref.\ \onlinecite{Tangri03}, where a partial differential equations is 
constructed that exhibits characteristics of \soc, and in which 
natural variables can be used.)

Here, we introduce a new \soc model in the form of a CA 
for 
ion temperature gradient (ITG) mode driven turbulence
in fusion plasma.
A main 
characteristic of the model is that it is constructed
in terms of  
the usual physical variables 
and that the temporal evolution of the CA,
which necessarily is in the form of rules, mimics actual 
physical processes as they are considered to be active in
the system.
All elements of the model 
are thus consistently interpretable in the usual physical  
way, no sand-grain or sand-pile analogy is used.

In the low $\beta$ core plasma of a tokamak, where the magnetic field 
dominates over pressure, 
turbulence is driven by two 
main electrostatic micro-instabilities, the 
ITG
driven modes and the Trapped Electron Modes (TEM; e.g.\ Ref.\ \onlinecite{Garbet04}).
In principal, the two instabilities can co-exist, 
in the case though we consider here, where ion heating dominates, 
the ITG modes become unstable and actually 
are the dominant instability.
In the strong turbulence regime then, the radial ion temperature  
profiles stay close to marginally stable, the 
gradients are stuck to their critical values, which is termed 
profile stiffness or profile consistency (e.g.\ Refs.\ \onlinecite{Garbet04}, 
\onlinecite{Ryter01}, \onlinecite{Stroth98}).
This plasma behaviour is very reminiscent of \soc,
and it is a main purpose of the application shown in this article 
to adequately model it.

From the CA and \soc modeling point of view, 
the necessary conditions for a system to be able to reach the 
state of
\soc 
are:
(i) there must be a driving process
which systematically increases the 'stress' in the system, (ii)
a threshold dependent instability must be defined in some way, 
and (iii) a relaxation process must set on if somewhere the
instability threshold is exceeded.
All these processes should preferably act only in a local neighbourhood,
and they must be formulated in the form of discrete evolution rules.
ITG mode driven turbulence
obviously meets the three 
prerequisites for \soc:
the system is systematically driven, namely heated, the ITG mode
instability is threshold dependent, and there is obviously 
a process that relaxes the instabilities, given the fact that
ITG mode driven turbulence reaches a saturated stationary state
(e.g.\ Ref.\ \onlinecite{Garbet04}).

Sect.\ II presents the model and explains how it is derived from 
usual physical processes in the plasma, Sect.\ III contains the 
results, and a discussion and the conclusions are given in 
Sect.\ IV and V, respectively.

\section{The model}

We consider a one dimensional grid along the direction of the minor 
radius $a$ in a toroidal confinement device, such as the tokamak, 
with the modeled domain being $[R-sa,R+sa]$, with $R$ the major radius 
and $s<1$, 
so
that only the core 
region is taken into account.

The basic scalar grid variable at the grid sites $x_i\in [R-sa,R+sa]$, for 
$i=1,\ldots,L$, is 
the local {\it ion temperature} $T_i\equiv T(x_i)$, considered as a positive real 
number
(\ie assuming also non-integer values, in contrast to some sand-pile models). 
The grid is assumed to be equi-spaced, with grid-spacing 
$\D x =  2sa / (L-1)$. 
The grid-spacing is not considered 'infinitesimal', as in the solution of 
differential equations, but it is finite, of the size of some smallest scale 
of interest, here of the size of a typical ion Larmor radius.

\subsection{Instability criterion}\label{ss:ic}

As is well known, see e.g.\ Refs.\ \onlinecite{KDBH95} and \onlinecite{DafniExp}, the ITG mode 
instability has a critical dependence 
on the normalized scale length $R/L_{T}$, where 
$L_{T}$ is the ion temperature-gradient scale-length,
$1/L_{T} := \vert \nabla T\vert /T$, in the sense that ITG mode instabilities
are triggered if the condition 
\begin{equation}
		R\frac{|\nabla T|}{T}
	>  \frac{R}{L_{\text{\scriptsize crit}}}
\label{inst1}
\end{equation}
is fulfilled

In order to use the instability criterion in the CA model, we 
have to determine the temperature gradient $\nabla T$
at the grid sites $x_i$. 
We thereto interpolate $T_i$ locally in the neighbourhood $(i-1, i, i+1)$ around
the central site $i$ with a second order polynomial, and we
differentiate the polynomial at the point $x_i$, which yields 
\begin{equation}\label{eq:delta}
  \d_i :=  \frac{1}{2\D x} \frac{T_{i+1} - T_{i-1}}{T_i}
\end{equation}
as an approximation to $\nabla T/T$
(and which is identical with the expression that a central difference scheme 
approximation for $\nabla T$ would yield).
After all, the grid site $i$ is considered to be unstable if
\begin{equation}\label{eq:ic}
R  |\d_i| 
> R/L_{\text{\scriptsize crit}}
\end{equation}

\subsection{Redistribution rules}\label{ss:rules}

We assume a normal diffusive process to be triggered around 
the grid site where an instability occurs, 
\ie\ we assume an evolution according to a simple Fokker-Planck equation
without advective term,
\begin{equation}
\frac{1}{2} n k_B \p_t T = -\p_x q ,
\label{eq:diffu}
\end{equation}
with $k_B$ Boltzmann's constant, $n$ the number density, and $q$ the 
classical heat flux,
\begin{equation}
q^{(cl)}(x) = - \chi n  k_B \p_x T ,  
\label{eq:qcl}
\end{equation}
where $\chi$ is the heat diffusivity.

\subsubsection{General form of the redistribution rules}

In the CA, the local diffusion process is formulated in terms 
of a set of {\it redistribution rules}, which describe how,
in the case of an instability, the temperature in the local 
neighborhood is redistributed,
the latter being defined here as consisting of the central unstable site, 
say $i$, and its two nearest neighbours, $i-1$ and $i+1$.
The general form of the redistribution rules can be written as
\begin{subequations}\label{eq:genRedis}
\begin{align}
 T_i^+ \ \  &= T_i \ \ \ + F_0(T_{i-1}, T_i, T_{i+1};\d_i, R/L_{\text{\scriptsize crit}}) ,\\
 T_{i\pm1}^+ &= T_{i\pm1} + F_\pm(T_{i-1}, T_i, T_{i+1};\d_i, R/L_{\text{\scriptsize crit}}) ,
 \end{align}
\end{subequations}
where the values after redistribution are denoted with a plus sign ($^+$) as
superscript,
and $F_0$, $F_\pm$ are the changes in temperature, here written with their
possible general functional dependence.
In the derivation of specific redistribution rules, we will take into account
actual physical processes and restrictions, 
and we will also make some assumptions.

\subsubsection{Energy conserving local diffusion}

We demand the redistribution rules to conserve the total temperature 
of the three involved grid-sites. From a physical point of view, 
the energy should be conserved, i.e.\ the product of the local temperature 
and the 
local density, so that we actually make here the assumption that
the density is constant, which can partly be justified by the 
small spatial extent of the local neighbourhood, which is 
of the order of three ion Larmor radii, 
and by the fact that the density in tokamaks usually is quite 
close to uniform. 
It is to note that 
the local conservation of the grid variable in the CA is an 
important property for the system to be able to reach the \soc state.
>From Eq.\ (\ref{eq:genRedis}),
the condition for local temperature conservation can be 
written as 
\begin{equation}
F_- + F_0 + F_+ = 0 . 
\label{eq:cons}
\end{equation}

Assuming the density to be constant in a local neighbourhood,
Eq.\ (\ref{eq:diffu}) and (\ref{eq:qcl}) reduce to a simple 
diffusion equation,
\begin{equation}
\frac{1}{2}  \p_t T =  \chi    \p_{xx} T  ,
\label{eq:diffus}
\end{equation}
where we have also assumed that the diffusivity $\chi$ is constant 
over a local neighbourhood. 
Diffusion 
according to Eq.~\eqref{eq:diffus}  results, under normal conditions,
in the local smoothing of the temperature profile,
and if the boundary conditions are fixed around the local 
neighbourhood, a linear asymptotic profile will be reached
(in the one dimensional case we treat here), where diffusion
stops since $\nabla^2 T =\partial_x^2 T = 0$.
The relaxation process in the CA should thus describe such a local
profile smoothing, expressed in terms of rules,
and for simplicity we assume here the smoothed profile to equal
the asymptotic one and thus to be linear.

The assumption for the relaxed profile to be linear together
with the assumption of temperature conservation already determine
the temperature $T_i^+$ after relaxation at the central site $i$ of an 
unstable region.
Since the relaxed profile is linear, $T_i^+$ equals the mean temperature
in the neighbourhood, 
\beq
T_i^+ = \frac{1}{2}(T_{i-1}^+ + T_{i+1}^+)  ,
\eeq
into which we insert Eq.\ (\ref{eq:genRedis}),
\beq
T_i + F_0 = \frac{1}{2}(T_{i-1} + F_- + T_{i+1} + F_+)  ,
\eeq
which we rearrange
\beq
2F_0 - F_- - F_+
=  -2T_i + T_{i-1} + T_{i+1}  .
\label{eq:h1}
\eeq
Now we note that $2F_0 - F_- - F_+ = 3F_0 -[ F_0 + F_- + F_+]$,
the sum in the square-bracket is though zero due to energy conservation
[Eq.\ (\ref{eq:cons})], and Eq.\ (\ref{eq:h1}) determines  
the increment at the central site
\beq
F_0 = -\frac{2}{3} \left( T_i - \frac{1}{2}\left(T_{i-1} + T_{i+1}\right)\right) .
\label{F01}
\eeq
For convenience, we define 
\beq
\tau_i := T_i - \frac{1}{2}\left(T_{i-1} + T_{i+1}\right) .
\label{taui}
\eeq

In the following, the increments $F_{\pm}$ of the nearest neighbours 
will be derived from a condition on the local heat fluxes.

\subsubsection{Heat fluxes}\label{ssec:heatf}

In the relaxation events described by Eq.\ (\ref{eq:genRedis}), heat is 
transported in the local neighbourhood, and
we quantify this
heat transport 
by determining 
the associated heat-fluxes. 
The starting point is the definition of the
dynamic heat flux,
\begin{equation}
q^{(dy)}(x) = \frac{1}{2} n(x) k_B T(x) \langle v(x) \rangle ,
\label{qdy}
\end{equation}
where $\langle v(x) \rangle$ is a local average velocity with which 
heat is flowing.

We consider a local neighbourhood $i-1, i, i+1$ around the grid site $i$,
which is unstable.
The mean location $s_i$ of the total heat content within the neighbourhood
can be defined by using the temperature values $T_j$ as weights,
\begin{equation}
s_i := \frac{\displaystyle\sum_{j = i-1}^{i+1} 
            j\Delta x\,T_j}{\displaystyle\sum_{j = i-1}^{i+1} T_j},
\label{eq:si}
\end{equation}
and where we have made use of the assumption of constant density.
Generally, $s_i$ will not coincide with a grid-point.
During the relaxation event, the local heat content will move to 
a new mean location $s_i^+$, so that the heat displacement is
$s_i^+ - s_i$. 
This displacement takes place in one time-step of duration $\Delta t$,  
so that the velocity with which heat is transported is 
\begin{equation}
	v_i := (s_i^+ - s_i)/\Delta t.
\label{eq:vi}
\end{equation}
According to Eq.\ (\ref{qdy}), we 
can now define the heat flux of a relaxation event as
\begin{equation}
	q_i := \frac{1}{2} n k_B v_i\sum_{j=i-1}^{i+1} T_j,
\label{eq:qi}
\end{equation}
or, with Eqs.\ (\ref{eq:vi}) and (\ref{eq:si}) and using the 
conservation of heat in relaxation events,
$\sum_{j=i-1}^{i+1} T_j=\sum_{j=i-1}^{i+1} T^+_j$, 
\begin{equation}
	q_i^{(dy)} := \frac{1}{2\Delta t} n k_B  
              \displaystyle\sum_{j = i-1}^{i+1} j\Delta x\,(T^+_j-T_j) .
\label{eq:qif}
\end{equation}
Inserting the redistribution rules in the form of Eqs.\ (\ref{eq:genRedis}) 
into Eq.\ (\ref{eq:qif}) yields
\beq
  q_i^{(dy)} 
      = \frac{\Delta x}{2\Delta t} 
                     n k_B \left[(i-1)F_- + iF_0 + (i+1)F_+\right],
\eeq
where the sum in the square brackets equals 
$i(F_- + F_0 + F_+)+F_+-F_-$, which, with Eq.\ (\ref{eq:cons}),
reduces to $F_+-F_-$, so that the heat flux finally writes as
\beq
  q_i^{(dy)} 
      = \frac{\Delta x}{2\Delta t} n k_B \left[F_+ - F_- \right].
\label{eq:qdy1}
\eeq

\subsubsection{Local diffusion of normal nature}

In order the local diffusion process to be of normal nature,
we demand that heat can only be transported from a hotter 
to a colder site, 
since else we would introduce anomalous diffusive effects
at the local scale. To ensure this condition, we demand that
the dynamic heat flux $q_i^{(dy)}$, Eq.\ (\ref{eq:qdy1}), equals the classical 
heat flux $q_i^{(cl)}$ of Fourier's law, Eq.\ (\ref{eq:qcl}).
In this way, heat is transported always in the direction of the
driving gradient ('downhill').

We discretize the gradient in Eq.\ (\ref{eq:qcl}) 
with the same method as used in the derivation of the instability criterion 
in Sect.\ \ref{ss:ic},
$$ \partial_x T \simeq \frac{T_{i+1} - T_{i-1}}{2\D x}, $$
so that 
$q^{(cl)}(x) = - \chi n  k_B \p_x T$   
is approximated as 
\begin{equation}
  q_i^{(cl)} = - \frac{1}{2\Delta x} \chi n k_B  (T_{i+1} - T_{i-1}).
\label{eq:qcl1}
\end{equation}
With Eqs.\ (\ref{eq:qdy1}) and (\ref{eq:qcl1}), the condition
\beq
q_i^{(dy)} = q_i^{(cl)}
\eeq
writes as 
\beq
    \frac{\Delta x}{2\Delta t} n k_B \left[F_+ - F_- \right]
= - \frac{1}{2\Delta x} \chi n k_B  (T_{i+1} - T_{i-1}),
\eeq
in which, due to energy conservation [Eq.\ (\ref{eq:cons})],
we can replace $-F_-$ by $F_0 + F_+$,
\beq
  \left[2F_+ + F_0 \right]
= - \frac{\Delta t}{(\Delta x)^2} \chi   (T_{i+1} - T_{i-1}),
\eeq
where $F_0$ is given by Eq.\ (\ref{F01}), so that the final form of the 
increment $F_+$ is 
\beq
F_+ =  
- \frac{1}{2} 
\left(F_0 + \frac{\Delta t}{(\Delta x)^2} \chi (T_{i+1} - T_{i-1})\right).
\label{F11}
\eeq
With Eq.\ (\ref{eq:cons}), the increment $F_-$ follows as $F_- = - F_0 - F_+$
from Eq.\ (\ref{F11}), 
\beq
F_- =  
- \frac{1}{2} 
\left(F_0 - \frac{\Delta t}{(\Delta x)^2} \chi (T_{i+1} - T_{i-1})\right),
\label{F22}
\eeq
and, for convenience, we define the normalized heat 
diffusivity \sig as 
\begin{equation}\label{eq:sig_chi}
  \sig \equiv \frac{\Delta t}{(\D x)^2} \, \chi .
\end{equation}

We note that we do have physical units for $\Delta x$, which equals
the ion Larmor radius, $\Delta t$ though is in arbitrary units, as
explained in Sect.\ \ref{algo} below, so that also $\sig$ necessarily is in
arbitrary units.


\begin{figure}[!ht]%
	\subfloat[Unstable state.]{
		\includegraphics[scale=0.7]{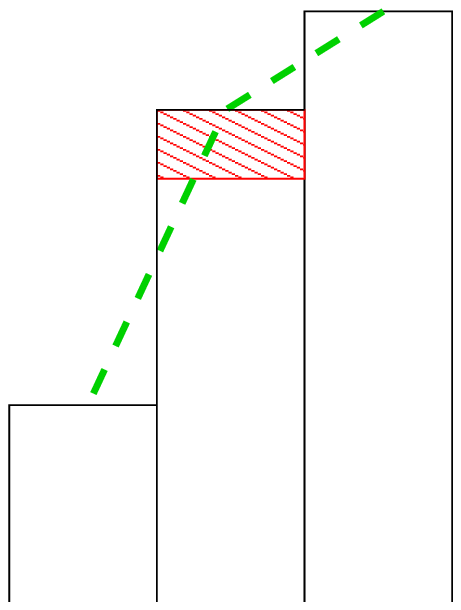}
		\label{fig:unstable}
	}\hfill
	\subfloat[Relaxed state.]{
		\includegraphics[scale=0.7]{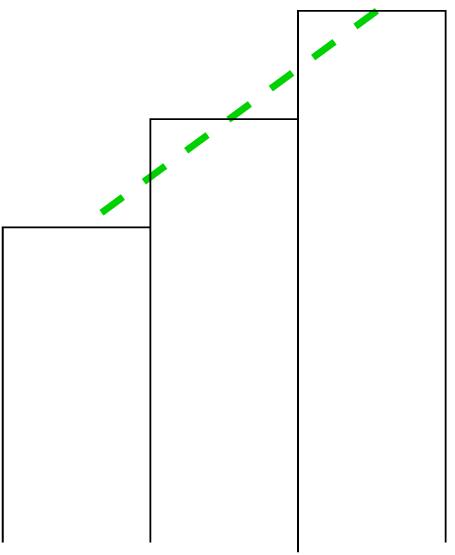}
		\label{fig:asterisk}
	}
	\caption{(Color online.) Sketch of the relaxation process. The red hatched region 
in the unstable configuration (a) marks the excess amount of heat to be moved, 
and the blue hatched region in the relaxed state (b) marks the amount of heat 
received in the relaxation event. The green 
dashed line connects the temperatures at the grid sites and represents
the local temperature profile.}
	\label{fig:relaxation}
\end{figure}

\subsubsection{Summary of the relaxation rules}

After all, the relaxation rules are (see Eqs.\ 
(\ref{eq:genRedis}), 
(\ref{F01}), 
(\ref{taui}), (\ref{F11}), (\ref{F22}), (\ref{eq:sig_chi}))
\begin{subequations}\label{eq:redis}
\begin{align}
	T_i^+ &= T_i - \frac{2}{3}\tau_i , \\
	T_{i\pm1}^+ &= T_{i\pm1} + \frac{1}{3}\tau_i \mp \frac{\sig}{2}(T_{i+1} - T_{i-1}) ,
\end{align}
\end{subequations}
with free parameter the normalized heat diffusivity 
$\sig >0$,
and where the increments $F_0$ and $F_\pm$  
in Eq.\ \eqref{eq:genRedis} are given as      
\begin{subequations}\label{eq:redisFuns}
\begin{align}
	F_0 &= -\frac{2}{3}\tau_i,\\
	F_\pm &= \frac{1}{3}\tau_i \mp \frac{\sig}{2}(T_{i+1} - T_{i-1}) .
                 \label{eq:redisFunsb} 
\end{align}
\end{subequations}
A sketch of the relaxation process is 
shown in Fig.\ \ref{fig:relaxation}.

The main effect of the local diffusion process 
in Eq.\ (\ref{eq:diffu}) should be to remove the cause of the diffusion,
i.e.\ instabilities should be relaxed, 
and we numerically find that the redistribution rules
we derived indeed relax the instabilities, i.e.\ it
holds that $R|\d_i^+| < R/L_{\text{\scriptsize crit}}$,
as long as the normalized diffusivity is positive and not too small,
$\sigma \agt 0.01$.

Last, we note that the redistribution rules in Eq.\ (\ref{eq:redis}), 
without though the parameter $\sigma$ (i.e.\ formally for $\sigma =0$), 
were used in Refs.\ \onlinecite{LH91}, \onlinecite{Isliker2000}, and 
\onlinecite{Isliker2001}
in the astrophysical context of Solar flares (in combination with an 
instability criterion different from Eq.\ (\ref{eq:ic})).

\subsection{\Bcs}\label{ssec:bcs}

We assume a constant value $T_b$ of the pedestal temperature 
outside the domain of the \ca.
$T_b$ is used in the calculation of the instability criterion, 
Eq.\ (\ref{eq:ic}), and in the redistribution rules,
Eq.\ (\ref{eq:redis}), at the edge of the system, 
formally entering the equations as $T_0=T_{L+1}=T_b$. 

The redistribution rules together with the initial conditions
(see below) ensure that there is no heat in-flow from the boundaries,
since always $T_b< T_1$ and $T_b < T_L$, the temperature
is always lower outside than inside the grid.

\subsection{Driving process}

The system undergoes a driving (heating) process, in which  
the temperature is increased, simulating the effect of heat injection. 
The loading process as such is not further specified, and it is 
understood to represent Ohmic 
heating as well as external heating. 

Concerning the temporal characteristics of the loading process,
we apply two variants, (i) we apply continuous loading, 
where the system is heated in every time-step (see the next section
for a definition of the time-step), 
and (ii) we consider loading only in stable configurations, where 
there are no instabilities present in the system. 

Moreover, we consider different spatial heating patterns:
global heating everywhere in the system, and heating only in
localized regions. The latter case includes heating only 
in the central region, reminiscent of Ohmic heating, heating 
only in a region off-axis, as in the case of rf heating, and a 
combination of central and off-axis heating, representing Ohmic
and off-axis heating applied simultaneously.

In either case of heating pattern, 
a grid site is chosen at random
from the region that undergoes heating, and 
the temperature at this site is increased by a constant, predefined 
amount $\Delta S_j$. 
The value of the parameter $\Delta S_j$ is chosen small 
enough to avoid over-driving the system, as usual for \soc models.

\subsection{Description of the algorithm\label{algo}}

The time evolution of the model is as follows:
(i) In a first step, the grid is scanned, and a list of the unstable sites
is created. (ii) If there are unstable sites, 
the instabilities are relaxed during a second scan of the grid.
Prior to redistribution, each site is checked again for 
stability and it is relaxed only if it is still found to be 
unstable.
The reason for checking again is the following:
It may happen that two neighbouring sites are simultaneously 
unstable, the 
redistribution rules are then first applied to just
one of them, 
whose
relaxation may 
cause the other site not to be unstable
anymore, and it would be physically unmotivated to redistribute its 
temperature.

If the system is loaded only in the absence of instabilities,
the steps (i) and (ii) are repeated until there are no
instabilities in the system anymore, and only then the system is loaded
again.
Each completion of steps (i) and (ii) is considered a time-step,
whose duration we consider to be $\Delta t=1$ in arbitrary units.
The repetition of steps (i) and (ii), from the first appearance of 
an instability until the system has reached an everywhere 
stable state, is termed an avalanche, as usual in the context of \soc 
models.
Continuous loading of the system means that 
after the completion of a time-step
the system
is also loaded, independent of whether there are instabilities present
in the system or not. In this case, there is no practical definition
of avalanches possible.

\section{Results\label{resultsr}}

\subsection{Parameters}

The parameters of the model are chosen to coincide with those
of the Joint European Torus\cite{JET} (JET) experiment.
We thus assume 
a minor radius $a = 1.25\,$m and a major radius $R=2.95\,$m,
and since the ITG instabilities are considered to dominate
turbulence only in the core, we restrict the spatial domain
used in the simulations to $\left[R-0.8 a \,,\,R+0.8 a\right]$.
The grid-size $\D x$ is  
of the same order as the ion Larmor radius $\rho_i$, so that, with 
$\rho_i \simeq 0.49\,$cm under typical conditions for JET, we use
a grid of $L = 401$ sites with grid-size $\D x = 0.5\,$cm.

According to Ref.\ \onlinecite{DafniExp}, 
the threshold $R/L_{\text{\scriptsize crit}}$ of 
the ITG mode instability can assume values in
the range $[3.5, 5]$.
In the following, 
and unless stated otherwise,
the instability threshold 
is set to $R/L_{\text{\scriptsize crit}} = 4$,
the normalized diffusivity to $\sigma = 0.5$,
the heating increment
to $\D S_j = 0.5$,
and we apply a constant value 
of the pedestal temperature 
$T_b = 500$ 
outside the domain of the model as boundary condition.
For the initial 
configuration we let the temperature everywhere equal 
$T_b$.

The basic free parameters of the model, 
which mainly determine the results and
which can be used 
to adjust the model to experimental data, 
are the threshold
$R/L_{\text{\scriptsize crit}}$, 
the pedestal temperature at the boundaries $T_b$,
and the normalized diffusivity $\sigma$.

\begin{figure}[!ht]
\includegraphics[width=0.4\textwidth]{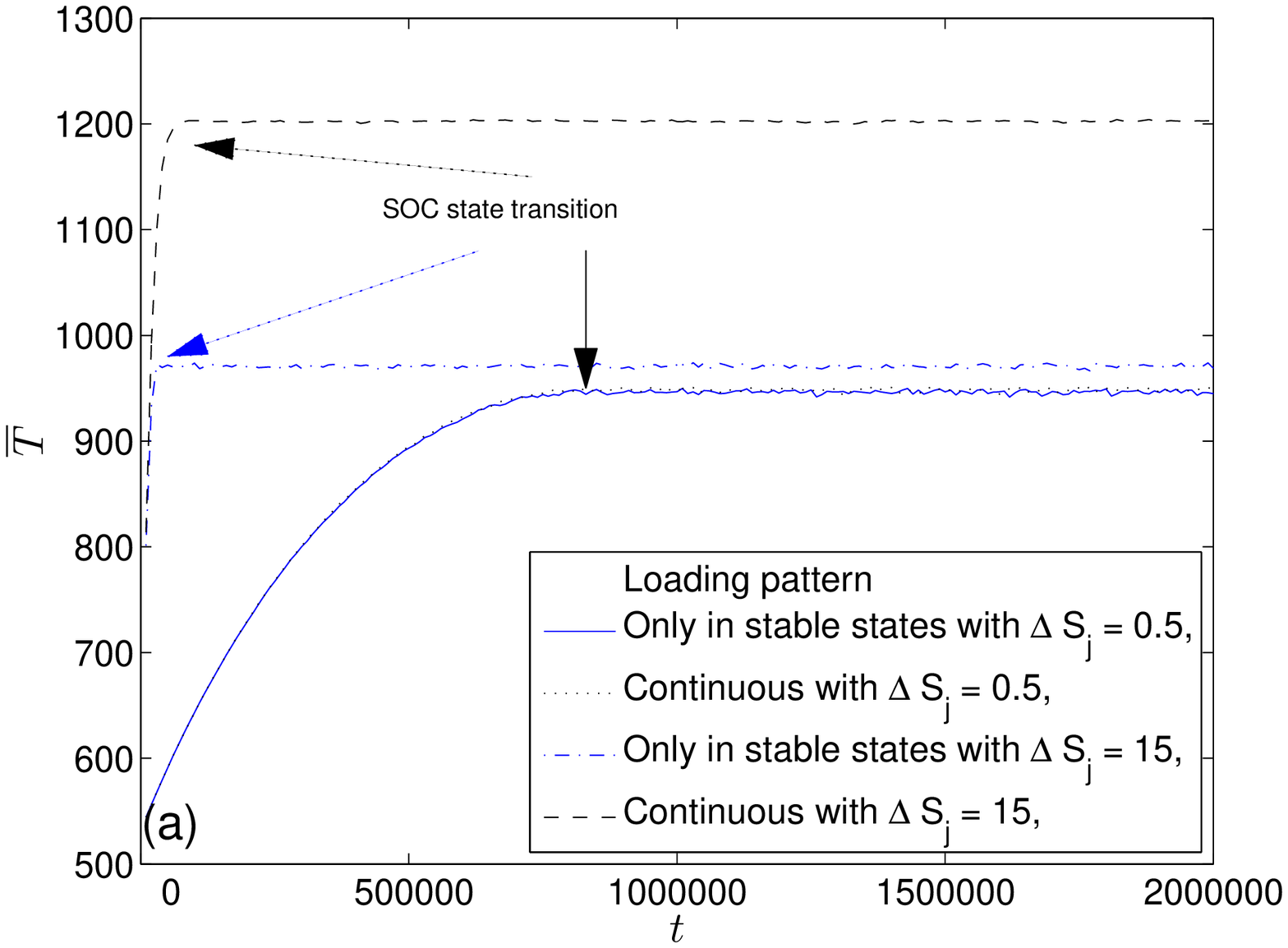}
\includegraphics[width=0.4\textwidth]{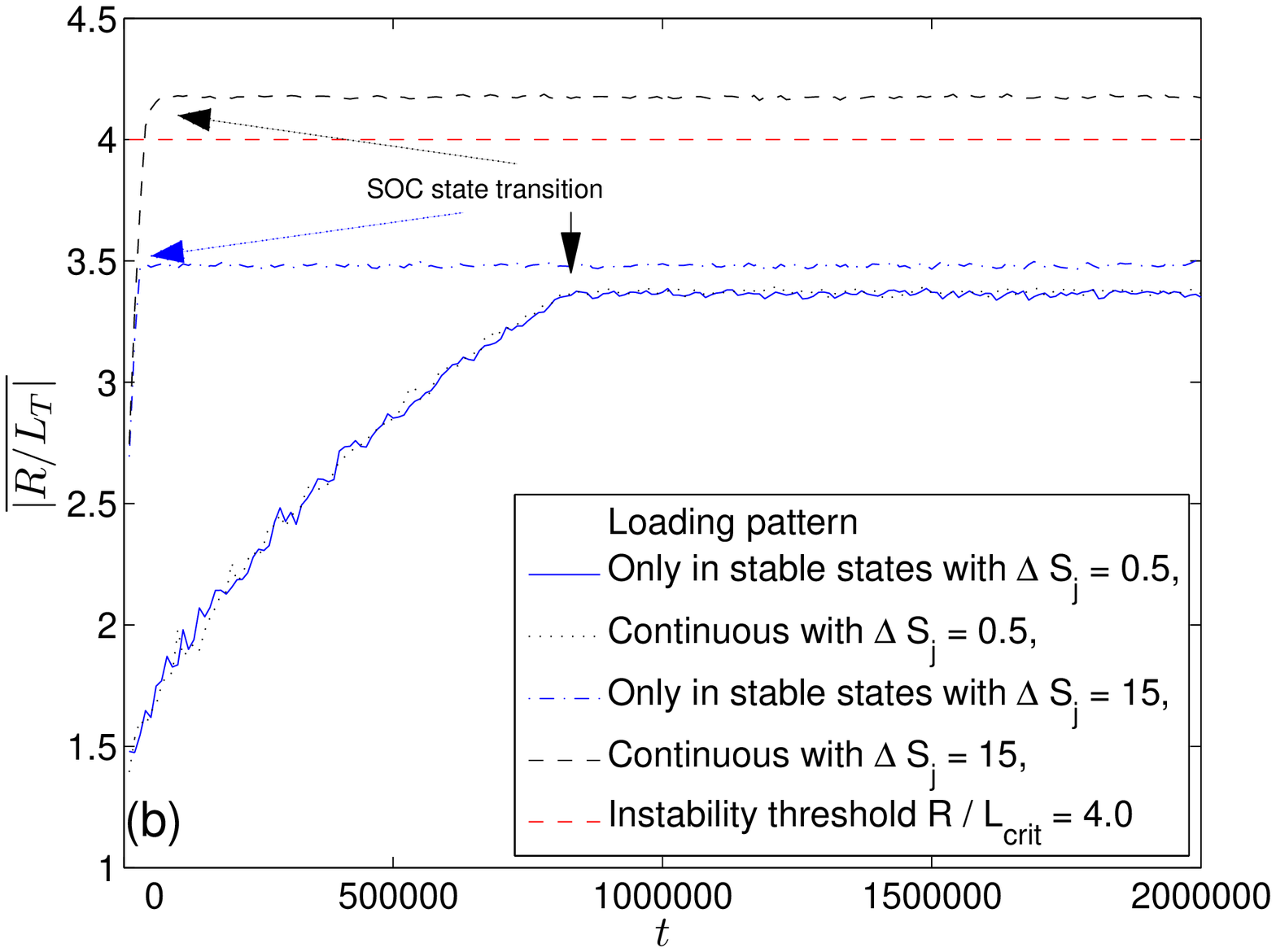}
\caption{(Color online.) 
Spatial mean of (a) the temperature \bar{T} and of (b)
the absolute normalized scale length
$\bar{|R/L_T|}$ 
as a function of time, 
for heating  only in stable states and for continuous 
heating, and with two different heating intensities.
The red 
horizontal dashed line in (b) shows the instability threshold 
$R/L_{\text{\scriptsize crit}}$, 
while the   
arrows mark the transition of the system to the \soc state.}
\label{fig:soc}
\end{figure}

\begin{figure}[!ht]
\includegraphics[width=0.4\textwidth]{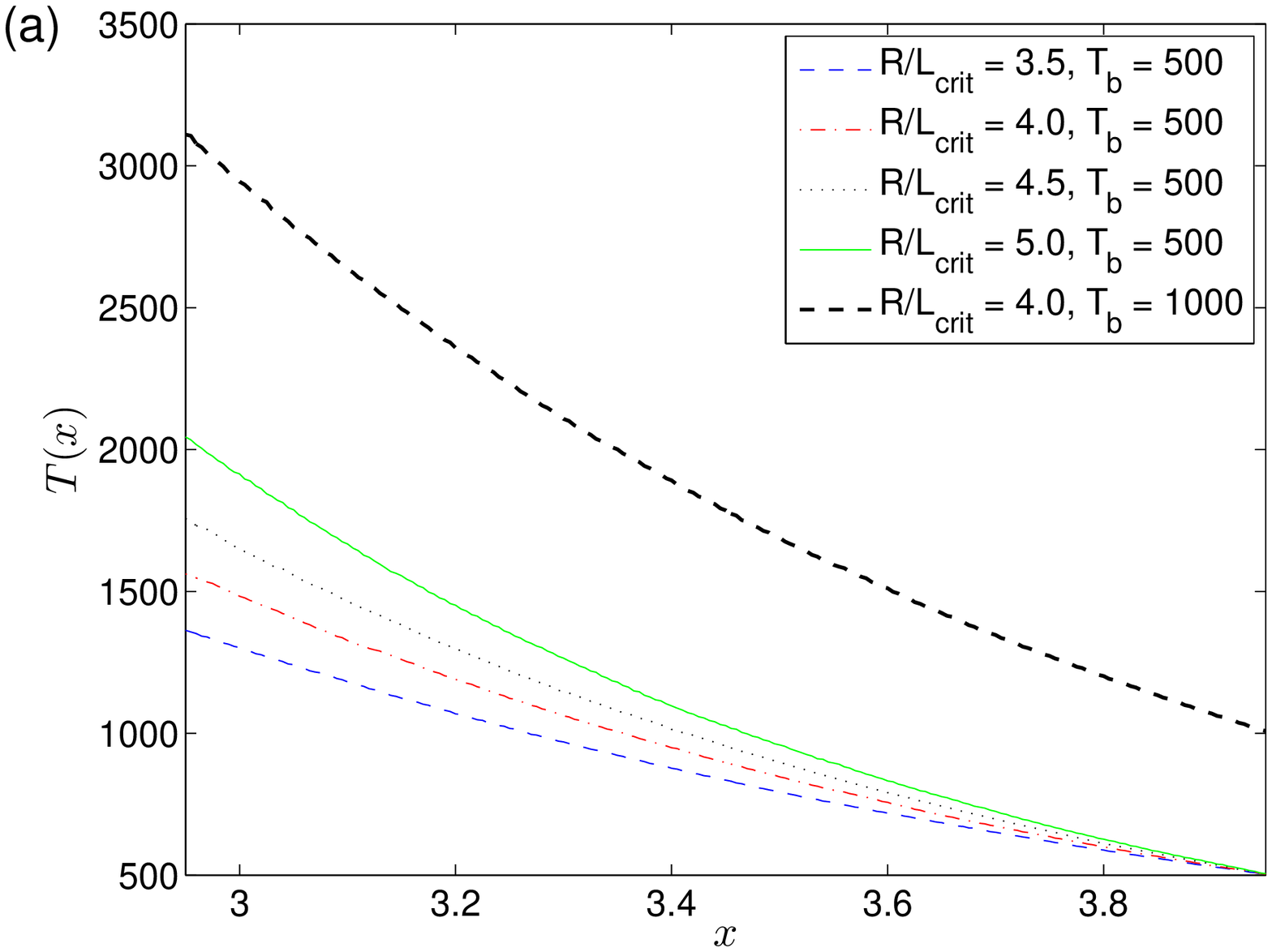}
\includegraphics[width=0.4\textwidth]{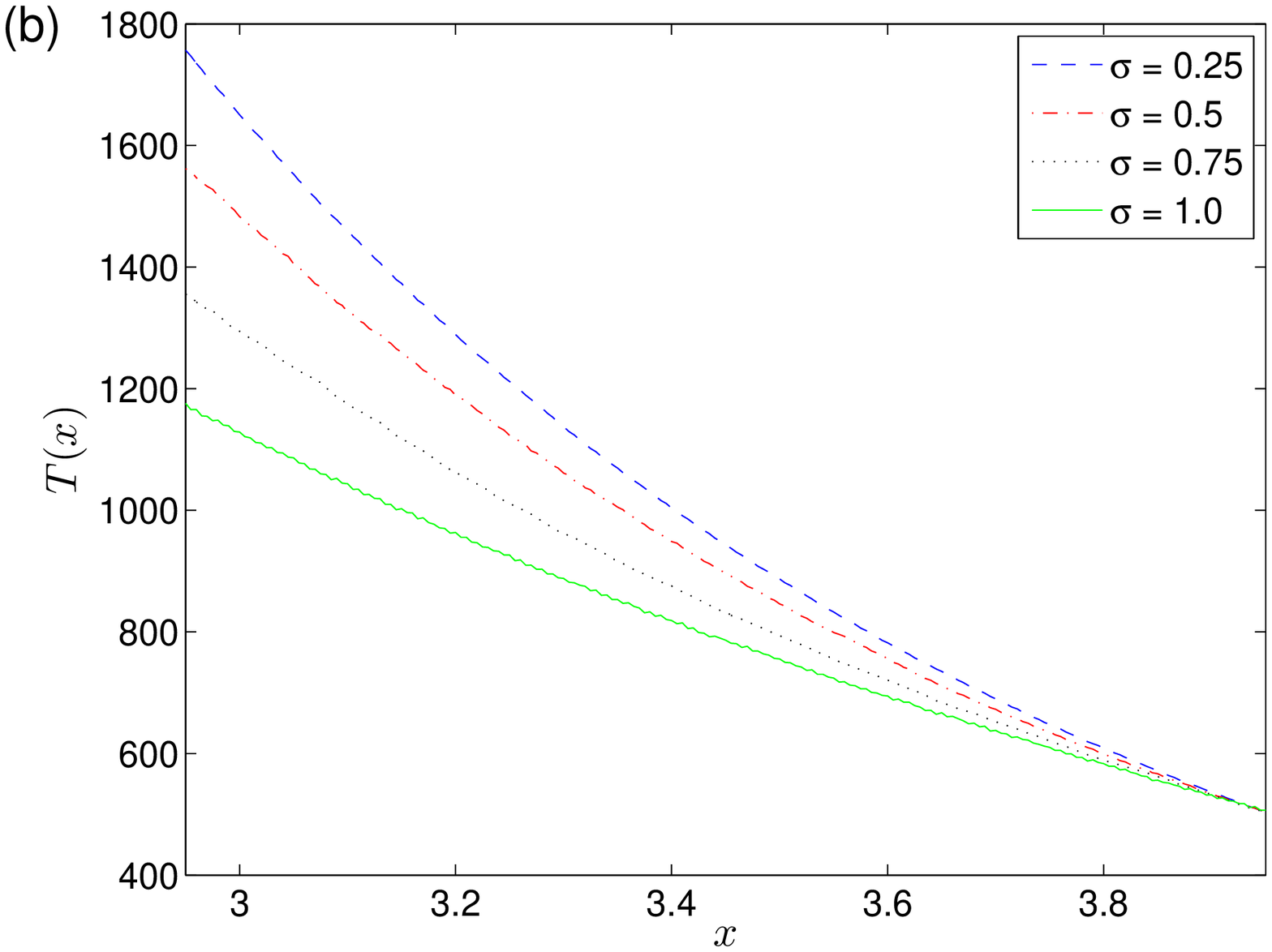}%
\caption{(Color online.) Temperature profiles during the \soc state for 
(a) different
values of 
$R/L_{\text{\scriptsize crit}}$
and 
$T_b$, 
and 
(b)
for different values of
\sig,
all with loading only in stable configurations.
}
\label{fig:final_conf}
\end{figure}

\subsection{The \soc state\label{socstate}}

Fig.~\ref{fig:soc} shows the spatial mean of the temperature $\bar{T}(t)$ 
and 
of the absolute 
local normalized scale length $\bar{|R/L_T|}$ 
for typical simulations with global heating,
one with loading only in stable configurations and
one with continuous loading, and for two different heating intensities. 
There is a transient phase, in which both mean values
are growing, and after which there is a turn-over to a
dynamic equilibrium state, where the mean values 
fluctuate around an asymptotic value. 
Once the asymptotic state is reached, avalanches
of widely varying sizes start to appear, from very small ones, bounded
to a local region, to avalanches that sweep through
the entire system. The reaching of an asymptotic value
of 
$\bar{|R/L_T|}$
and 
$\bar{T}(t)$, 
together with the appearance
of avalanches of all sizes, are  
indicative for the system to have reached the \soc state.
The duration of 
the transient phase depends on the intensity of the 
driving process. 
It is to note that, in the case of intense continuous loading, the 
mean normalized scale length can also be above critical, whereas in the 
case of loading only in stable configurations it is always below critical.

Temperature profiles in the \soc 
state are shown in Fig.~\ref{fig:final_conf}(a) for 
different values of the threshold 
$R/L_{\text{\scriptsize crit}}$ and pedestal temperature $T_b$
(here and in the following, the right halves of the symmetric 
profiles are shown).
In all cases, 
the profiles are of exponential shape, and the threshold, 
together with the value $T_b$ of the pedestal temperature at the boundaries,
determine the maximum value that is reached in the center:
the central temperature increases with increasing threshold as well as
with increasing pedestal temperature, the latter moreover 
causing a shift of the 
entire temperature profile towards higher values. 

During the 
\soc state, the dynamically evolving temperature profiles stay very close to 
the characteristic exponential shape, even the distortions caused by  
large avalanches would hardly be visible in the scales of 
Fig.~\ref{fig:final_conf}. This implies that the total energy content of 
the system is, within small fluctuations, 
constant, heat is ejected from the system at the boundaries at the
same statistical rate it is injected by heating.

In 
Fig.~\ref{fig:final_conf}(b), 
we show the temperature profiles during \soc state 
for different values of the normalized diffusivity \sig, 
keeping the pedestal temperature and the 
threshold fixed.
The central peak temperature decreases with increasing diffusivity,
a large $\sigma$ causes heat to be faster redistributed in the system 
and also to be more efficiently ejected at the edge.

\subsection{Central and off-axis heating\label{coh}}

\begin{figure}[!ht]
\includegraphics[width=0.4\textwidth]{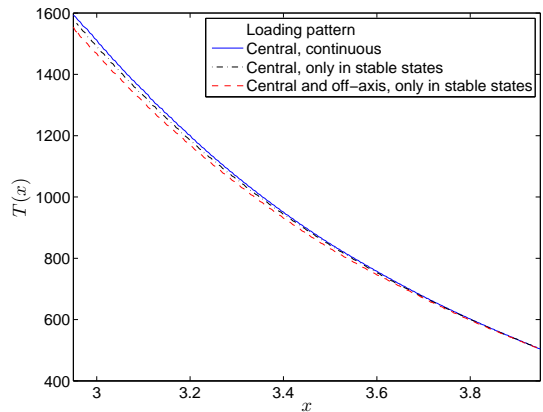}
\caption{(Color online.) Temperature profiles for central heating (black dash-dotted), 
and for simultaneous central and off-axis heating (red dashed),
both with heating only in stable states,
and for central, continuous heating (blue, solid).
\label{fig:conf_c-oa}}
\end{figure}

In order to investigate the impact of off-axis heating on the
temperature profile, we apply 
two 
spatial heating patterns, 
one for pure central heating and one for simultaneous central heating and
equally strong off-axis heating
(the central region being defined as the grid-sites $i\in [151,251]$,
and the off-axis region as the sites $i\in [351,401]$).  
In Fig.~\ref{fig:conf_c-oa},  the temperature profiles 
yielded by the two simulations
are shown during the \soc state. 
Also the profile from off-axis heating is clearly peaked at the center, 
there is no sign of the off-axis heat source visible, 
and the difference between the two profiles is actually very small.
This implies that the temperature profiles
exhibit a very high degree of stiffness, they are basically not
influenced by the applied spatial heating pattern. This also
implies that there is strong anomalous transport, where heat
is transported against the driving gradient ('uphill'),
despite the fact that diffusion at the local 
level is imposed to be of a normal kind.

\subsection{Continuous loading}

In the cases of temperature profiles shown so-far, 
heating was applied only when there 
were no instabilities present in the system. This is justified
physically only if the ITG mode instabilities together with their
relaxation have a much faster 
time-scale than the heating process. When the two time-scales are
comparable, loading should take place more frequently.

In Fig.~\ref{fig:conf_c-oa}, we also show a temperature profile
for continuous heating
in the central region, here heating
in every time-step. The profile is also of exponential shape, and it is 
similar to the one with heating only in stable states, 
the temperature values are though slightly 
larger,
because there is more heat injected into the system in this case,
and the system has less time to eject heat at the edges (see
also Fig.\ \ref{fig:soc}). In the case
of continuous heating, the value of the central peak temperature  
depends also on the heating increments $\Delta S_j$, 
it increases with increasing $\Delta S_j$,
whereas in the case of loading only 
in stable configurations it is almost independent of $\Delta S_j$.

\begin{figure*}[!ht]
\includegraphics[width=0.4\textwidth]{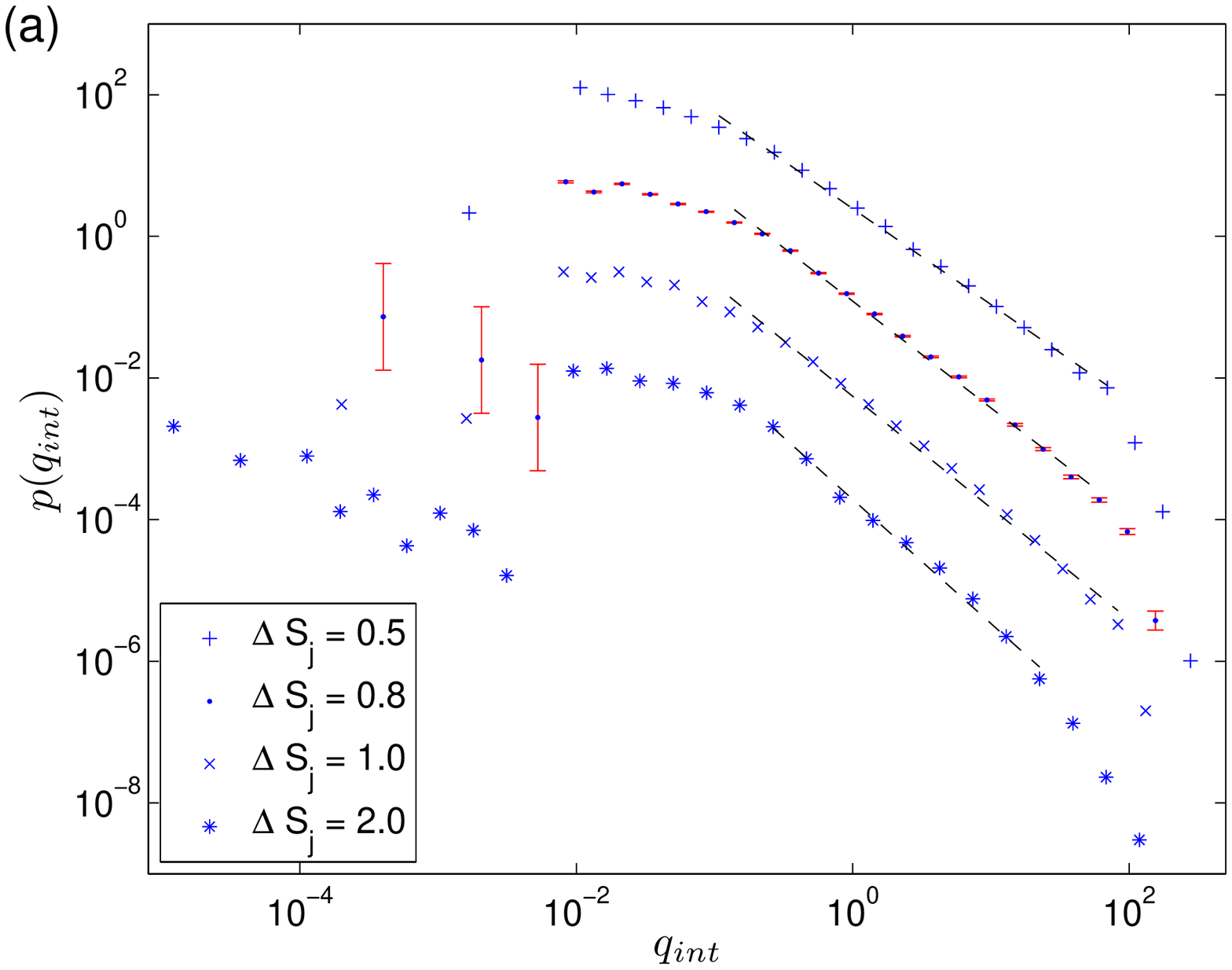}
\includegraphics[width=0.4\textwidth]{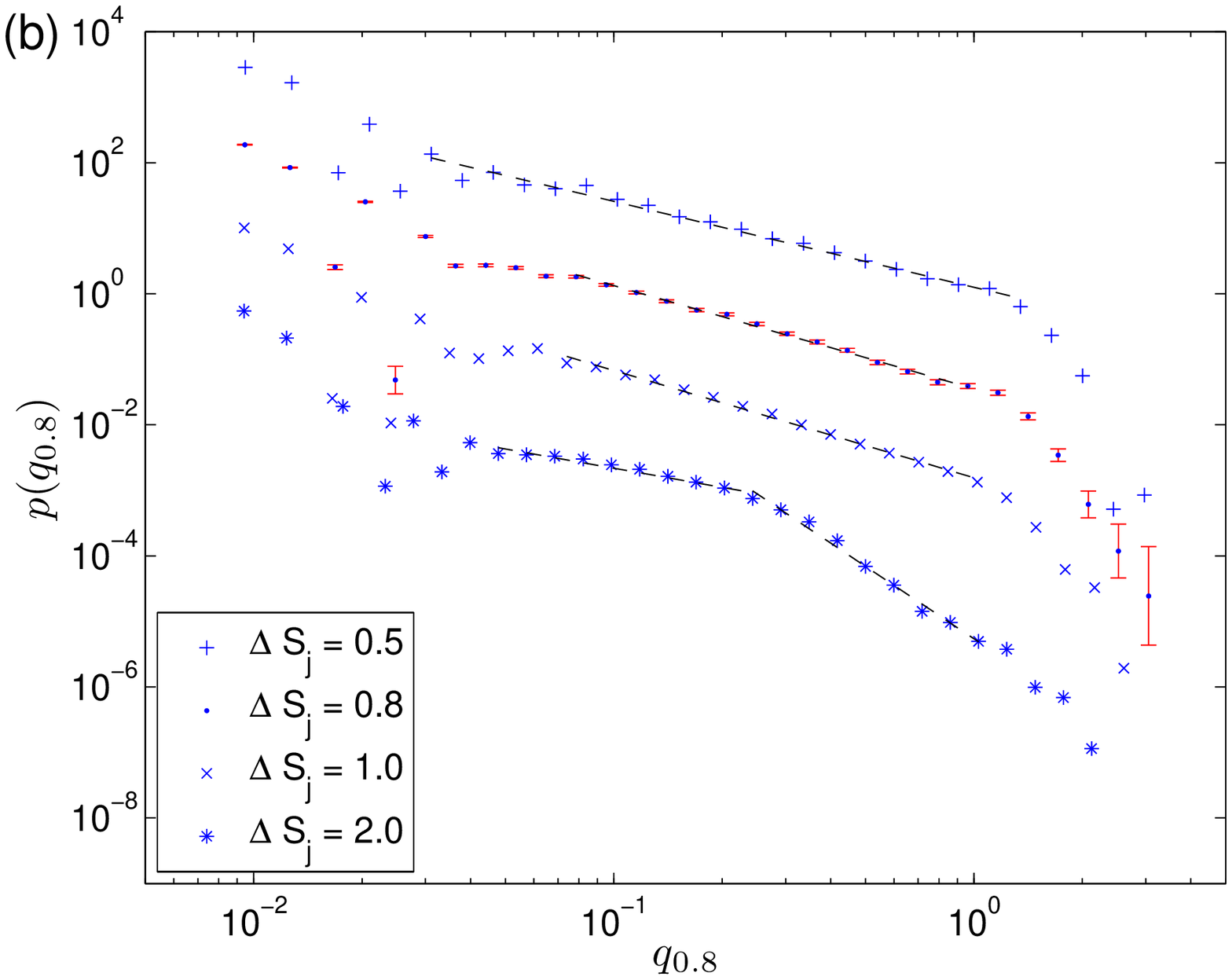}
\includegraphics[width=0.4\textwidth]{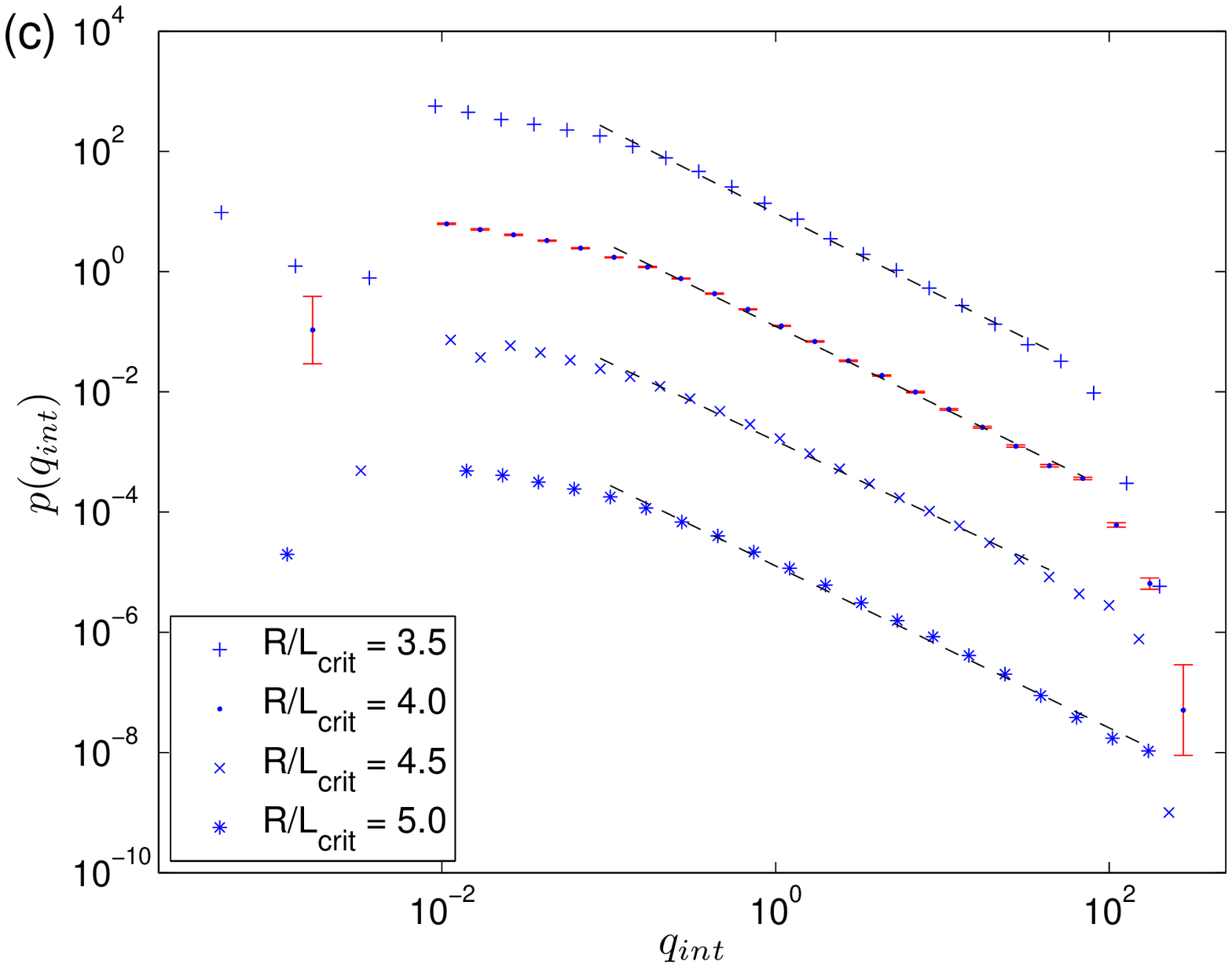}
\includegraphics[width=0.4\textwidth]{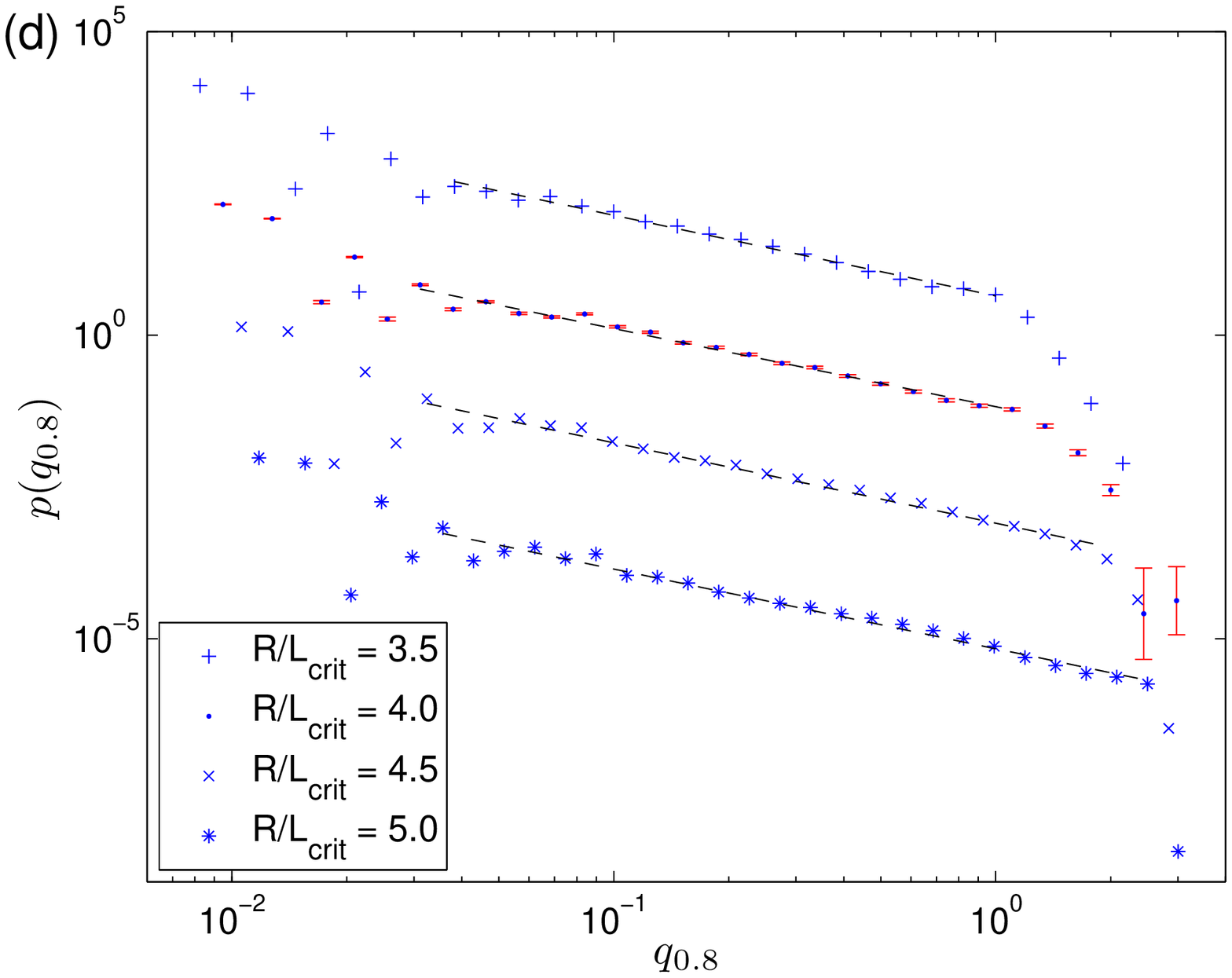}
\includegraphics[width=0.4\textwidth]{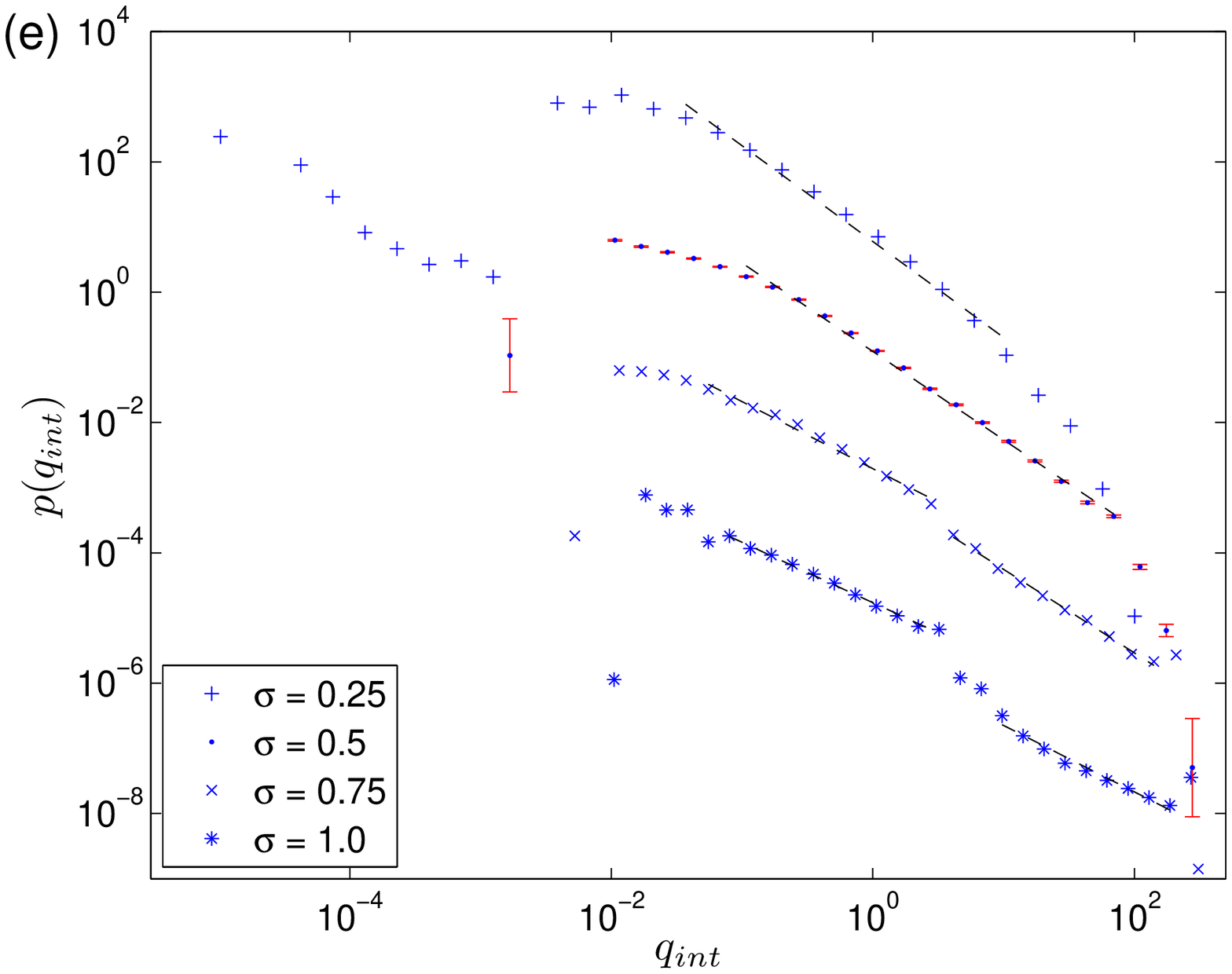}
\includegraphics[width=0.4\textwidth]{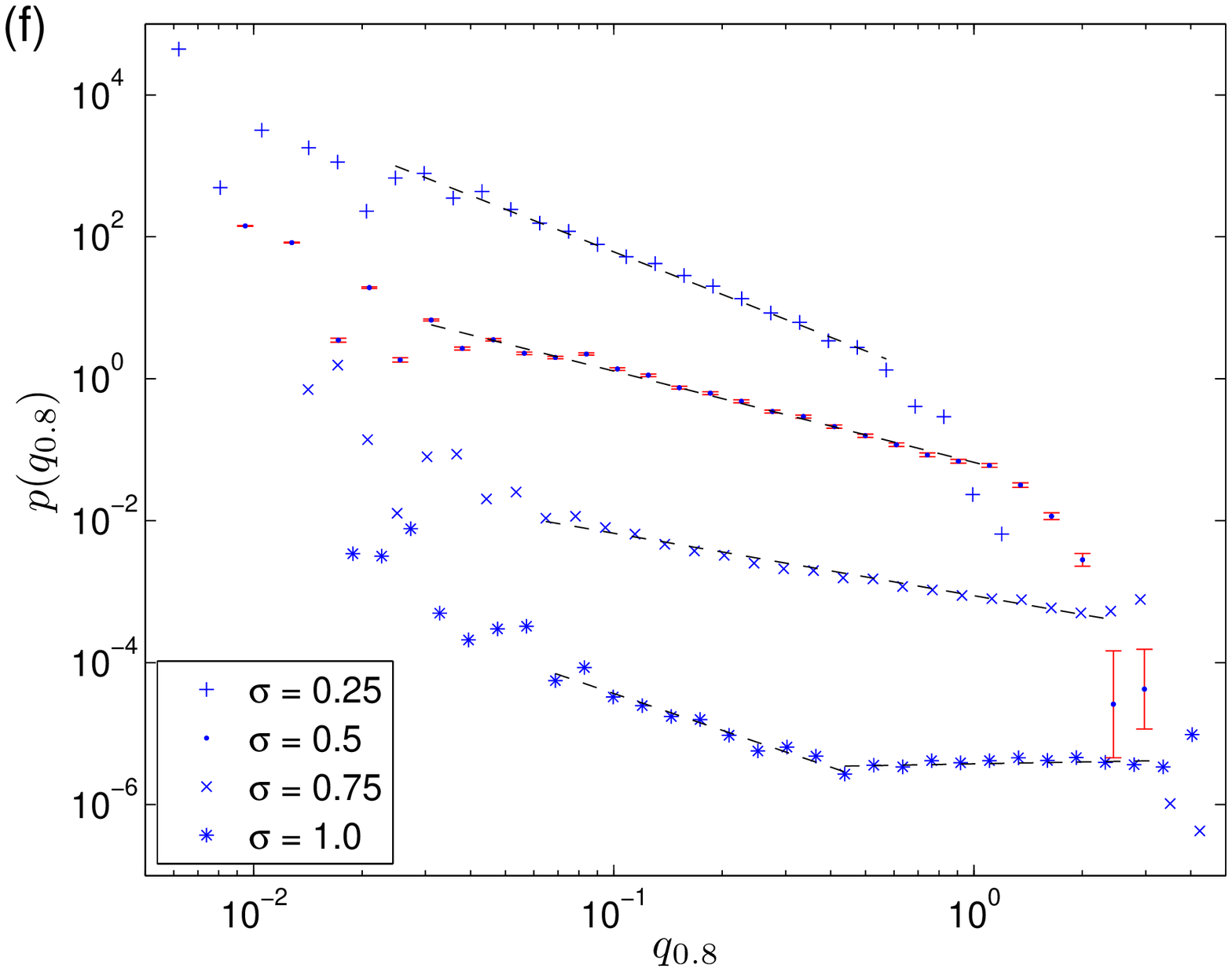}
\caption{(Color online.) Distributions (normalized histograms) of the total heat fluxes 
per avalanche $q_{int}$ [(a), (c), (e)]  
and total heat-outfluxes per avalanche $q_{0.8}$ [(b), (d), (f)], 
for different loading intensities $\Delta S_j$ [(a), (b)],
for different thresholds $R/L_{\text{\scriptsize crit}}$ [(c), (d)],
and for different diffusivities $\sigma$ [(e), (f)],
together with power-law fits (dashed lines).
(The distributions are shifted in the vertical direction for better
visualization, and in each sub-figure error-bars (red) are shown for reference 
in one case.)
}
\label{fig:outflow_c}
\end{figure*}

\begin{table*}[!ht]
\caption{Power-law indices $\alpha$ and $\beta$ from the power-law fits
to the distributions of the total internal fluxes $q_{int}$ and the 
total out-fluxes $q_{0.8}$, respectively, in Fig.\ \ref{fig:outflow_c}.
\label{tabind}}
\begin{tabular}{c r r}
\hline\hline
$\Delta S_j$ & \ $\alpha$ & $\beta$ \\
\hline
0.5 & 1.36 &   1.30 \\
0.8 & 1.51 &   1.57 \\
1.0 & 1.58 &   1.63 \\
2.0 & \ 1.75 & \ \ 0.97 \& 3.67  \\
\hline\hline
\end{tabular}
\begin{tabular}{c r r}
\hline\hline
$R/L_{\text{\scriptsize crit}}$ & \ $\alpha$ & $\beta$ \\
\hline
 3.5  &   1.39 & 1.30 \\
 4.0  &   1.36 & 1.30 \\
 4.5  &   1.30 & 1.30 \\
 5.0  & \ 1.35 & 1.30 \\
\hline\hline
\end{tabular}
\begin{tabular}{c r r}
\hline\hline
$\sigma$ & \ $\alpha$ & $\beta$\\
\hline
0.25 & 1.47           & 2.00 \\
0.50 & 1.36           & 1.30 \\
0.75 & 1.03 \& 1.28   & 0.90 \\
1.00 & \ 0.93 \& 1.02 & \ \ 1.70 \& -0.90 \\
\hline\hline
\end{tabular}
\end{table*}

\subsection{Distribution of heat out-fluxes\label{disho}}

The local heat flux per 
relaxation event is given by
$q_i^{(dy)}$ in Eq. (\ref{eq:qdy1}), 
so that the total heat flux in a time-step (see Sect.\ \ref{algo}) 
is 
\begin{equation}
	Q_{t} := \sum_{i=1}^{L} q_i(t) ,
\label{eq:qts}
\end{equation}
and the total internal heat flux of an avalanche $q_{int}$ follows as
\begin{equation}
	q_{int} := \sum_{t\, \in \,\textrm{avalanche}} Q_{t} .
\label{eq:qav}
\end{equation}
Correspondingly, the total heat out-flux per avalanche at the right edge 
$q_{0.8}$ 
 is given  
by setting $i=L$ in Eq. (\ref{eq:qdy1}) (i.e.\ considering the radial 
position $x=0.8a$) and summing over the avalanche,
\begin{equation}
	q_{0.8} := \sum_{t \in avalanche} q_{L}(t) .
\label{eq:qLL}
\end{equation}

For the 
case of global heating in stable configurations,
Fig.\ \ref{fig:outflow_c}
shows the normalized distributions of the 
internal fluxes $q_{int}$ and 
the heat 
out-fluxes $q_{0.8}$ from a large number of avalanches, 
and for different values of the parameters $\Delta S_j$, 
$R/L_{\text{\scriptsize crit}}$, and $\sigma$.
All the distributions are of
clear single or double power-law shapes in the intermediate range,
extending over roughly 3 decades for $q_{int}$ and 1.5 decades for
$q_{0.8}$. The larger extent of the power-laws of $q_{int}$ reflects 
the fact that, from its definition,
$q_{int}$ has a larger dynamic range. 
The power-laws we find, together with Fig.\ \ref{fig:soc}, 
are indicative for the system to be in the state of \soc.

The indices $\alpha$ and $\beta$ of the fitted power-laws 
($p(q_{int}) \propto q_{int}^{-\alpha}$ and 
$p(q_{0.8}) \propto q_{0.8}^{-\beta}$, respectively)
are summarized in Table \ref{tabind}:
(1) The power-law indices are basically independent of the threshold
$R/L_{\text{\scriptsize crit}}$. (2) They depend on the heating 
intensity $\Delta S_j$, the dependence is though very weak if the 
$\Delta S_j$ are small, i.e.\ for low-level heating,  
the latter being defined in the sense
that the system, when in \soc state and just having become stable, 
must on the average 
be heated with 
much more than just one heat increment $\Delta S_j$ in order to get 
unstable again.
(3) The power-law indices clearly depend on the diffusivity, they 
decrease and the distributions become flatter with increasing $\sigma$, 
a high diffusivity at the local level thus favours large heat-fluxes
at the macroscopic scale, as one also would intuitively expect.

\section{Discussion}

\subsection{Marginally stable profiles\label{msp}}

>From Eq.\ (\ref{inst1}), we can analytically determine the marginally stable 
configuration, defined through
$|\nabla T(x)|/T(x) = 1/L_{\text{\scriptsize crit}}$, which has the general solution
$  T(x) = c e^{- |x|/L_{\text{\scriptsize crit}}}$,
and where $c$ depends on the boundary conditions we apply,
$T(x=\ell)=T_b$ (with $\ell:= 0.8a$), from where 
$c = T_b e^{\ell/L_{\text{\scriptsize crit}}}$, so that
\begin{equation}\label{eq:exactSol}
  T(x) = T_b e^{-(|x|-\ell)/L_{\text{\scriptsize crit}}}.
\end{equation}
(An exponential, marginally stable profile is also derived
in Ref.\ \onlinecite{Garbet04b}, in the frame of a critical gradient
transport model.)
The dynamic profile of the model in the \soc state 
follows the shape of the marginally stable profile,
with a smaller decay constant though, i.e.\ reaching lower
temperature and thus being sub-marginal when the system is loaded
only during stable states, and 
with a
larger than critical decay constant 
in the
case of intense enough continuous loading (see Fig.~\ref{fig:soc}).
The maximum temperature in the centre is 
$T(x=0) = T_b e^{\ell/L_{\text{\scriptsize crit}}}$
according to Eq.\ (\ref{eq:exactSol}),
and in the numerical simulations we indeed find the peak temperature
to be proportional to $T_b$ and $1/L_{\text{\scriptsize crit}}$ 
(see Fig.\ \ref{fig:final_conf}(a)), and to be dependent on $\ell$ such that 
larger systems reach larger temperatures in the center.

It is worthwhile noting that 
Eq.\ (\ref{eq:exactSol}) does not give a complete qualitative description 
of the system dynamics, it does not contain any information 
on the local relaxation process and therewith on the third free parameter, 
the normalized diffusivity $\sigma$. The latter has though a decisive 
influence on e.g.\ the central peak temperature reached, as  
Fig. \ref{fig:final_conf} shows, the peak temperature increases
with decreasing $\sigma$, and the profile actually approaches the 
marginally stable profile for very small $\sigma$.

\subsection{Comparison with experimental data}

Experimental ion temperature profiles in tokamaks are well known to be very 
stiff and of exponential shape in the inner core region
(see e.g.\ Ref.\ \onlinecite{Garbet04b}).
Examples of published, exponentially shaped ion temperature profiles 
include different devices and confinement modes, e.g.\ 
the L-mode in Tokamak Fusion Test Reactor, TFTR\cite{TFTR}  
(Fig.\ 1 in Ref.\ \onlinecite{KDBH95}), 
the H-mode in ASDEX Upgrade\cite{ASDEXupgrade} (Fig.\ 2 in Ref.\ \onlinecite{Peeters02}), 
the L-mode in ASDEX Upgrade (Fig.\ 2 in Ref.\ \onlinecite{Wolf03}), 
and the L-mode in DIII-D\cite{DIII_D} tokamak (Fig. 1(c) in Ref.\ \onlinecite{Baker03}),
and the high degree of stiffness in experimental ion temperature 
profiles is presented and analyzed in e.g.\ Ref.\ \onlinecite{DafniExp}.
Moreover, Ref.\ \onlinecite{Rhodes99} experimentally finds 
power-law distributions
for particle fluxes (with power-law index $1$), which can be related
to heat fluxes under the assumption that 
the heat-flux is convective.

We thus can conclude that generally a good qualitative agreement of the model 
with experiments can be 
achieved with respect to 
(i) the exponential profile shape (Sect.\ \ref{socstate}),
(ii) the high profile stiffness (Sect.\ \ref{coh}),
and to some degree there is also a qualitative agreement with respect to
(iii) the power-law distribution of out-fluxes (Sect.\ \ref{disho}). 
Last, we note that, having used parameters of the JET experiment
(concerning the system size, ion Larmor radius, and pedestal 
temperature $T_b$) and assuming $T_b$ to be in units of eV, 
(iv) we naturally find a dynamic range of the ion temperature 
that is comparable to the one experimentally seen at JET, 
see e.g.\ Fig.\ 2(e) in Ref.\ \onlinecite{DafniExp} in comparison 
with our Fig.\ \ref{fig:final_conf}.
For a quantitative comparison, we would need to find a way to 
calibrate the time and therewith the normalized diffusivity in the 
model.

\subsection{Remarks on the SOC modeling approach}


The purpose of the study presented here was to introduce a \soc model for the 
ion temperature dynamics in 
the core region of tokamaks, with the elements of the model being physically 
interpretable in a sound and consistent way. This implies that in the 
derivation of 
the instability criterion, relaxation rules, and loading process, we had to  
be guided 
by and take into account the basic physical variables and processes that are 
active 
in the system (to the degree that they are known). 

The \soc model was constructed from four basic pieces of 
information:
(i) the form of the instability criterion, Eq.\ (\ref{inst1}), which was taken
from 
experiments 
on ITG mode driven turbulence (Fig.\ 1 in Ref.\ \onlinecite{DafniExp}), 
(ii) a simplified Fokker Planck equation from general transport theory, 
Eqs.\ (\ref{eq:diffu}) and (\ref{eq:diffus}), 
(iii) the definition of the classical (Fourier's law, 
Eq.\ (\ref{eq:qcl})) and dynamical (Eq.\ (\ref{qdy})) heat flux, and
(iv) the concept of \soc in its operational definition (basically following 
Ref.\ \onlinecite{Bak87}, and also Ref.\ \onlinecite{LH91}). 

The operational definition of \soc actually expresses the necessary conditions 
for an extended system to reach the \soc state, which partly were 
already mentioned in the Introduction. 
An extended system disposed to \soc must 
(a) systematically be driven, 
(b) allow local, threshold-dependent instabilities, 
(c) posses a mechanism that relaxes the instabilities locally, and 
(d) the system variable considered must be conserved in the relaxation events.
In the model construction, the concept of \soc played the guiding role, 
in that it determines the elements of the model and their form, 
it contains the formal prerequisites for the final model to be a realization of 
the \soc concept. 

In the classical sand-pile model of Ref.\ \onlinecite{Bak87} 
and in many of its descendants mentioned in the introduction, 
instabilities occur if the height difference between neighbouring sand-columns
exceeds a threshold, where the height difference can be interpreted 
as a very rough approximation 
to the gradient $\nabla h$ of the local sand-column height $h$.
In the tokamak core plasma though, instabilities occur when $\nabla T/T$
exceeds a threshold (see Sect.\ \ref{ss:ic}),
not $\nabla T$. The core plasma thus belongs to a formally different class of 
unstable systems than the sand-pile, which implies that we 
cannot make use of the sand-pile paradigm in core plasma \soc modeling,
as long as it is our explicite aim to incorporate the actual physics of the 
problem under consideration as close as possible into the \soc model.

It is worthwhile noting that \soc models that belong to different 
instability classes
exhibit different characteristic spatial profiles of the grid variable. 
The sand-pile models, with an instability criterion formulated in terms of 
an approximation to the gradient $\nabla h$ of the sand-column height, exhibit linear profiles, 
see Ref.\ \onlinecite{Bak87}.
In an astrophysical context, 
Refs.\ \onlinecite{LH91}, \onlinecite{Isliker2000}, and 
\onlinecite{Isliker2001}
use an approximation to the Laplacian, $\nabla^2 \mathbf{A}$,  
in the instability criterion, 
with $\mathbf{A}$ the vector potential of the magnetic field,
and they find parabolic spatial profiles for the components 
of $\mathbf{A}$. The instability criterion
used here, an approximation to $\nabla T/T$, leads to exponential 
spatial profiles, as shown in Sect.\ \ref{resultsr}.
The reason for these characteristic differences is that, 
in all cases, the spatial profiles in \soc state  
follow the marginally stable profiles in shape, 
which are different for the different classes of instabilities.

Last, we note that our model has a high computational 
efficiency, which 
is a great advantage over e.g.\ the gyro-kinetic approach, 
it allows for fast interpretation and analysis of experimental results, 
and, after some 
further development, it even might be of potential use for plasma control.
For example, the temperature profiles can easily be modeled globally, 
in the entire 
core region, heating patterns of any kind can easily be applied and explored, 
and the statistics of the heat fluxes can be determined very accurately, since  
the model can be monitored over very long times.

\section{Conclusion}

We introduced a \ca model that implements  
the basic physics of ITG driven turbulence, as relevant for the core of fusion 
plasmas in tokamak devices, in L-mode as well as in H-mode. 
The model is formulated in terms
of evolution rules, which makes it computationally very efficient.
The rules implement two processes, heating by simple heat
deposition, and local diffusion if a threshold in the normalized 
ion temperature gradient $R/L_T$ is exceeded, whereby it is ensured
that energy is conserved and that heat is never transported from a cooler to a 
hotter site.
The model is formulated in terms of   
the usual physical variables, also in what the threshold is concerned, 
and the actual physical processes are directly mimicked in terms of rules.

The model also represents an implementation of \soc, which it
always reaches after an 
initial transient phase. In the \soc state then, 
the model yields symmetric ion temperature profiles of exponential 
shape. These profiles exhibit very high stiffness, in that 
they basically are independent of the loading pattern applied
(central and off-axis heating yield the same profile). 
This implies that there is anomalous 
heat transport ('uphill' heat transport, against the driving gradient) 
present in the system, despite the fact that diffusion at the local 
level 
is imposed to be of a normal kind.

In a qualitative comparison of the model's basic properties
with experimental data, 
we find good 
agreement,  
at least for time instances
where the experimental profiles also 
exhibit an exponential shape. We thus can conclude that the 
physical system we investigate, i.e.\ ITG driven turbulence, 
is qualitatively compatible with the \soc state of our model.

\begin{acknowledgments}
Work performed under the Contract of Association 
Euratom-Hellenic Republic. It is the sole responsibility of the authors and 
does not necessarily represent the views or opinions of any of the sponsors, 
which do not bear any responsibility for its contents.
H.I. is grateful to D.\ Carati, J.J.\ Rasmussen, and B. Weyssow 
for helpful discussions.
\end{acknowledgments}

\cleardoublepage
\end{document}